\newcommand{\inkr}{ \in \mskip -14mu / \mskip 9mu}
\documentstyle[11pt]{article}

\title{NON-ABELIAN FLUX ALGEBRAS IN YANG-MILLS THEORIES 
\thanks{To appear in Proceedings of XXVIII  ICHEP'96.}}
\author{Leszek \L ukaszuk}
\date{Soltan Institute for Nuclear Studies, \\
Ho\.{z}a 69, 00-681 Warsaw,
Poland \\
e-mail: lukaszuk@fuw.edu.pl}

\begin{document}
\maketitle
\begin{abstract}
Contour gauges are discussed in the framework of
canonical formalism. We find flux operator algebras with
the structure constants of underlying Yang-Mills theory.
\end{abstract}
\maketitle

\newpage
\section{Introduction}
Contour gauges \cite{ll1} have been already successfully applied 
to Yang-Mills theories in the framework of path integral approach.
In this note I am going to discuss these gauges in the framework 
of canonical formalism.
The curves admitted will be those discussed
in ref.\cite{l1}, where a slight modification of the gauge condition
used in \cite{ll1} has been proposed and discussed under the name
of ponderomotive gauges. These gauges can be indexed \cite{l1}
with homotopy
families admissible by geometry of region $V$ considered. 
In the case of Yang-Mills theory we limit ourselves to self-contractible
families, defined in ref.\cite{l1}. They have a useful property,
namely Y-M potentials are orthogonal to these curves, i.e. from
the gauge constraints
\begin{equation}
\int_{c(x,x_{0})} A_{a}^{\mu}(y) dy_{\mu} = 0
\label{j1}
\end{equation}
follows \cite{l1}
\begin{equation}
\frac{\partial c_{\mu}(x, x_{0}, \tau)}{\partial \tau} A_{a}^{\mu}
(c(x, x_{0}, \tau)) = 0
\label{j2}
\end{equation}
for any $0 \leq \tau \leq 1$ and $x \in V$.\\
We are going to implement these gauges into canonical
formalism of Y-M theory (Ch.2). Next, we establish in Ch.3 algebras
of fluxes:
\begin{equation}
{\cal B}_{a}^{(\sigma)} = \int_{\sigma} B_{a}^{k} n^{k}
d\sigma
\label{j21}
\end{equation}
\begin{eqnarray}
{\cal E}_{a}^{(\sigma^{*})} = \int_{\sigma^{*}} 
E_{a}^{k} n^{k} d
\sigma 
\label{j3} \\
\vec{x}^{0} \inkr \sigma  \nonumber
\end{eqnarray}

\section{Dirac brackets for Y-M theory}
\setcounter{equation}{0}
In what follows the discussion of surface terms will be omitted.
The canonical Hamiltonian is then:
\begin{equation}
H = \int_{V} d^{3}x {\cal H}
\label{d1}
\end{equation}
with
\begin{eqnarray}
{\cal H} = \frac{1}{2} (\vec{B}_{a} \cdot \vec{B}_{a} + \vec{E}_{a}
\cdot \vec{E}_{a}) - 
[\vec{\nabla} \cdot \vec{E}_{a} - g {\cal C}_{abc} \vec{A}_{b} \cdot
\vec{E}_{c}] A_{a}^{0}
\label{d2}
\end{eqnarray}
\vspace{0.25cm}
\begin{equation}
E_{a}^{i}(\vec{x}) = - \Pi_{a}^{i}(\vec{x})
\label{d3}
\end{equation}
\begin{equation}
D_{a}^{(1)} = \Pi_{a}^{0} \approx 0
\label{d4}
\end{equation}
\begin{equation}
D_{a}^{(2)} = \vec{\nabla} \cdot \vec{E}_{a} - g 
{\cal C}_{abc} \vec{A}_{b} \cdot \vec{E}_{c} 
\approx 0
\label{d5}
\end{equation}
We take temporal gauge
\begin{equation}
D_{a}^{(3)} = A_{a}^{0} \approx 0
\label{d6}
\end{equation}
and ponderomotive space-like gauge constraint 
\begin{equation}
D_{a}^{(4)} = \int_{c(\vec{x}, \vec{x}_{0})} A_{a}^{i}(y)
dy^{i} \approx 0
\label{d7}
\end{equation}
The constraints (\ref{d4}-\ref{d7}) are compatible with
\begin{equation}
{\cal H}' = \frac{1}{2}(\vec{B}_{a} \cdot \vec{B}_{a} + \vec{E}_{a}
\cdot \vec{E}_{a}) + D_{a}^{(2)}(x) v_{(a)}^{(2)}(x)
\label{d8} 
\end{equation}
where
\begin{equation}
v_{a}^{(2)}(x) = \int_{c(\vec{x}, \vec{x}_{0})} dy^{i}
E_{a}^{i}(y)
\label{d9}
\end{equation}
Let us notice that the condition (\ref{d7}) is trivial for
$\vec{x}=\vec{x}_{0}$ (for self-contractible curves $\vec{c}(
\vec{x}_{0}, \vec{x}_{0}) = \vec{x}_{0})$. Therefore our further
considerations will be valid for the region $V_{-}$:
\begin{equation}
V_{-} = V - P(\vec{x}_{0})
\label{d10}
\end{equation}
Next, let us remark that compatibility of (\ref{d7}) with (\ref{d8})
is evident once we prove - in analogy with Maxwell theory \cite{l1}- that 
in $V_{-}$
\begin{equation}
[D_{d}^{(4)}(\vec{x}), D_{a}^{(2)}(\vec{y})]_{P} =
\delta(\vec{x} - \vec{y})
\delta_{ad}
\label{d11} 
\end{equation}
The first term of $D_{a}^{(2)}(y)$, $- \vec{\nabla}\cdot \vec{E}_{a}$
(comp. eq.(\ref{d5}))yields already r.h.s. of (\ref{d11}) - derivation
is the same as for Maxwell theory (comp.\cite{l1}). So we have to
show that
\begin{equation}
{\cal C}_{abc} [D_{a}^{(4)}(\vec{x}), A_{b}^{i} E_{c}^{i}(y)]_{P}
\approx 0
\label{d12}
\end{equation}
The use of (\ref{d7}) gives
\begin{eqnarray}
{\cal C}_{abc} [D_{a}^{(4)}(x), A_{b}^{i}(y) E_{c}^{i}(y)]_{P}= \nonumber \\
= {\cal C}_{abc} \int_{c(\vec{x}, \vec{x}_{0})} dz^{k} [A_{d}^{k}(z),
A_{b}^{i}(y)E_{c}^{i}(y)]_{P}= \nonumber \\
= {\cal C}_{abc} (-) \delta_{cd} \int_{c(\vec{x}, \vec{x}_{0})} dz^{k}
\delta(\vec{z}-\vec{y}) A_{b}^{k}(z)
\label{d13}
\end{eqnarray}
Please notice, that $dz^{k} A^{k}(z)|_{z\in c(\vec{x}, \vec{x}_{0})}
\approx 0$ from (\ref{d7}) (comp.eqns (\ref{j1}), (\ref{j2})),
therefore (\ref{d12}) is proved.\\
Let us come back to constraints ((\ref{d4})-(\ref{d7})). With the
help of (\ref{d11})the matrix
\begin{equation}
d_{a, b}^{i, k} = [D_{a}^{(i)}(x), D_{b}^{(k)}(y)]_{P}
\label{d14}
\end{equation}
can be written as
\begin{equation}
d_{a, b}^{i, k}=\left[ \begin{array}{cc}
0 & -I \\
I &  0
\end{array}
\right]^{ik} \cdot \delta_{ab} \delta_{3}(\vec{x} - \vec{y})
\label{d15}
\end{equation}
so that
\begin{equation}
d^{-1} = - d
\label{d16}
\end{equation}
and Dirac brackets of the theory follow \cite{l2}:
\begin{eqnarray}
[E_{c}^{r}(\vec{x}), A_{d}^{s}(\vec{y})]_{D} = \delta_{cd} \delta_{rs}
\delta(\vec{x}-\vec{y}) - \nonumber \\
- \left[ \frac{\partial}{\partial y^{s}} \delta_{cd} 
- g {\cal C}_{cdb} A_{b}^{s}(y) \right] 
\cdot \int_{w \in c(\vec{y}, \vec{x}_{0})} dw^{r} \delta(\vec{x}-\vec{w}) 
\label{d17}
\end{eqnarray}
\vspace{0.35cm}
\begin{equation}
[A_{c}^{r}, A_{d}^{s}]_{D} = 0
\label{d18}
\end{equation}
\vspace{0.27cm}
\begin{eqnarray}
[E_{c}^{r}(\vec{x}), E_{d}^{s}(\vec{y})]_{D} = \nonumber \\ 
= g {\cal C}_{cdf}
\left[ \int_{w\in c(\vec{x}, \vec{x}_{0})} dw^{s} \delta(\vec{y} -
\vec{w}) E_{f}^{r}(\vec{x}) 
+ \int_{w \in c(\vec{y}, \vec{x}_{0})} dw^{r} \delta(\vec{x} -
\vec{w}) E_{f}^{s}(y) \right] 
\label{d19}
\end{eqnarray}
In the next section eqns.(\ref{d17} - \ref{d19}) will be used in the derivation
of nonabelian algebras.

\section{Flux algebras}
\setcounter{equation}{0}
Let us consider at the beginning a special type of surfaces 
appearing in definitions (\ref{j21}), (\ref{j3})
of ${\cal B}$, ${\cal E}$
fluxes. Take a loop $L$ and some homotopy $c(\vec{x}, \vec{a})$.
We define a horn $H(L, c)$:
\begin{equation}
\vec{x} \in H(L, c) \Longleftrightarrow x^{k} = c^{k}(\vec{L}(t),
\vec{a}, t_{1})
\label{t11}
\end{equation}
for some $t, t_{1} \in [0, 1]$ and fix orientation on this surface:
\begin{equation}
\vec{n} || \left( \frac{\partial \vec{c}}{\partial t_{1}} 
\times \frac{\partial \vec{c}} 
{\partial t} \right)
\label{t1111}
\end{equation}
We are going to show that fluxes ${\cal B}, {\cal E}$ through these
homotopy horns are equal to loop integrals:
\begin{equation}
\int_{H(L, c)} (\vec{B}_{a} \cdot \vec{n}) d \sigma =
\int_{L} f_{a}^{r} d x^{r} 
\label{t12}
\end{equation}
\begin{equation}
\int_{H(L^{*}, c^{*})}(\vec{E}_{a} \cdot \vec{n}) d \sigma =
\int_{L} .^{*}f_{a}^{r} dx^{r}
\label{t13}
\end{equation}
with
\begin{equation}
f_{a}^{r}(x) = \int_{c} B_{a}^{k} \varepsilon^{kij}
\frac{\partial y^{j}}{\partial x^{r}} dy^{i}
\label{t14}
\end{equation}
\begin{equation}
^{*}f_{a}^{r}(x) = \int_{c^{*}} E_{a}^{k} \varepsilon^{kij}
\frac{\partial y^{j}}{\partial x^{r}} dy^{i}
\label{t15}
\end{equation}
$L, L^{*}$ and $c, c^{*}$ need not be related. At this stage we need
not specify in what gauge $\vec{B}_{a}, \vec{E}_{a}$ are given.
Eqns.(\ref{t12}), (\ref{t13}) are consequence of a simple observation.
Take any antisymmetric tensor $T_{ij}$ and define
\begin{equation}
g^{r}(x) = \int_{y \in c(\vec{x}, \vec{a})} T^{ij}(y)
\frac{\partial y^{j}}{\partial x^{r}} dy^{i}
\label{t16}
\end{equation}
Then
\begin{equation}
\int_{L} g^{r}(x) dx^{r} = \int dt dt_{1} 
\frac{\partial y^{i}}{\partial t_{1}}   
\frac{\partial y^{j}}{\partial t}
\varepsilon^{ijk} T^{k}(y)
\label{t17}
\end{equation}
where
\begin{equation}
T^{ij} = \varepsilon^{ijk} T^{k}
\label{t18}
\end{equation}
\begin{equation}
y^{i} = c^{i}(L(t), x_{0}, t_{1}) 
\label{t19}
\end{equation}
\begin{equation}
\int_{L} g^{r}(x) dx^{r} =
\int_{H(L, c)} (\vec{T} \cdot \vec{n}) d \sigma 
\label{t110}
\end{equation}
Replacement $g \longrightarrow f or^{*}f$
and $T^{k} \longrightarrow B^{k} or E^{k}$, gives eqns.(\ref{t12}) and 
(\ref{t13}), respectively.\\
If we specify $\vec{B}_{a}$, $\vec{E}_{a}$ to be in 
a gauge defined through $c$ from eqn.(\ref{t14}), then $f_{a}^{r}(x)$
is a potential in this gauge (comp.ref.\cite{l1}). Still there is
a vast choice of homotopies $c^{*}$ defining dual potential
$,^{*}f$. Let us consider Dirac brackets of ${\cal E}, B$
in c-gauge:
\begin{eqnarray}
\left[ {\cal B}_{d}, {\cal E}_{c} \right]_{D} =
\left[ \int_{L} f_{d}^{r} dx^{r}, \int_{L^{*}}.^{*}f_{c}^{s} dy^{s}
\right]_{D} = \nonumber \\
= \left[ \int_{L} f_{d}^{r} dx^{r}, \int_{L^{*}} dy^{s} 
\int_{c^{*}(y, a^{*})}dz^{i} \frac{\partial z^{j}}{\partial y^{s}} 
\varepsilon^{ijk} E_{c}^{k}(z) \right]_{D}
\label{t111}
\end{eqnarray}
Using (\ref{d17}) one gets, after some algebra, the following
expression:
\begin{eqnarray}
\left[ {\cal E}_{c}^{H^{*}}, {\cal B}_{d}^{H} \right]_{D} =
\delta_{cd} N(L; H^{*}) + \nonumber \\
+ g c_{cdg} \int_{x \in L} dx^{r} f_{g}^{r}(x)
N(c(x,a); H^{*})
\label{t3}
\end{eqnarray}
where
\begin{equation}
N(L; H^{*}) = \sum_{t} sgn \left( \frac{\partial \vec{L}(t_{1})}{\partial t_{1}}
\cdot \vec{n}_{H^{*}}(t_{2}, t_{3}) \right)
\label{t4}
\end{equation}
\begin{eqnarray}
N(c(\vec{x}, a); H^{*}) = 
\sum_{\tau(x)} sgn \left( \frac{ \partial \vec{c}(
\vec{x}, \vec{a}, \tau_{1})}{\partial \tau_{1}} \cdot \vec{n}_{H^{*}}(\tau_{2},
\tau_{3}) \right)
\label{t5}
\end{eqnarray}
with $t_{i}$, $\tau_{i}(x)$ being the solutions of the following equations:
\begin{equation}
\vec{c}^{*}(L^{*}(t_{2}), \vec{a}^{*}, t_{3}) = \vec{L}(t_{1})
\label{t6}
\end{equation}
\begin{equation}
\vec{c}^{*}(L^{*}(\tau_{2}), \vec{a}^{*}, \tau_{3}) = \vec{c}(\vec{x}, \vec{a}, 
\tau_{1})
\label{t7}
\end{equation}
and $\vec{n}_{H^{*}}$ being normal to a horn $ H^{*} \equiv H(L^{*}, c^{*})$
(comp.eqns (\ref{t11}), (\ref{t1111})).\\
The conditions (\ref{t6}) or (\ref{t7}) are fulfilled whenever the surface of
$H^{*}$ is pierced by loop $L$ or homotopy curve $c(\vec{x}, \vec{a})$, 
respectively. $N$'s in eqns (\ref{t4}), (\ref{t5}) denote net numbers of 
piercings.\\
The abelian part of (\ref{t3}) has been already discussed \cite{l3} for the
radial gauge; it leads to t'Hooft algebra \cite{l4}. The non-abelian part can be
expressed through surface integrals. Call $K_{N}$ part of a loop $L$,
characterized by $N(c; H^{*}) = N$, $N$ fixed ($K_{N}$ can consist of disjoint
pieces). We have $L = \sum_{N} K_{N}$ and corresponding horn surface:
\begin{equation}
H(L, c) = \bigcup_{N} H(K_{N}, c)
\label{t8}
\end{equation}
where
\begin{equation}
x \in H(K_{N}, c) \Longleftrightarrow \vec{x} = \vec{c}(\vec{y}, \vec{a}, t)
\label{t9}
\end{equation}
for some $Y \in K_{N}$ and $t \in [0, 1]$. Evidently eqn(\ref{t12}) holds for
$H(K_{N}, c)$ so that eqn(\ref{t3}) can be rewritten as:
\begin{eqnarray}
[{\cal E}_{c}^{H^{*}}, {\cal B}_{d}^{H}] = \delta_{cd} N(L; H^{*}) + 
\nonumber \\
+ g c_{cdg} \sum_{N} N \int_{H(K_{N}, c)} \vec{B}_{g} \cdot \vec{n} d \sigma
\label{t20}
\end{eqnarray}
Let us add, that in fact eqn(\ref{t20}) holds for any surface $S$, not 
necessarily a horn $H^{*}(L^{*}, c^{*})$. $H^{*}$ is useful if we want to keep
relation with loop integrals over dual potential (see eqns (\ref{t13}), 
(\ref{t15})). More generally, we have:
\begin{eqnarray}
[{\cal E}_{c}^{S}, {\cal B}_{d}^{H}] = \delta_{cd} N(L; S) 
+ g c_{cdf} \sum_{N} N \int_{H(K_{N}, c)} \vec{\cal{B}}_{f} \cdot \vec{n} 
d \sigma
\label{t21}
\end{eqnarray}
Let us consider now fluxes ${\cal E}_{c}^{(S_{1})}$, ${\cal E}_{c}^{(S_{2})}$.
Surfaces $S_{i}$ are parametrized by given $s_{i}(t_{1}, t_{2})$:
\begin{equation}
x \in S_{i} \Longleftrightarrow \vec{x} = \vec{s}_{i}(t_{1}, t_{2})
\label{t22}
\end{equation}
for some $(t_{1}, t_{2}) \in [0, 1]$.\\

The Dirac bracket of ${\cal E}_{c}^{(S_{1})}$, ${\cal E}_{d}^{(S_{2})}$ - 
calculated in $c$-gauge - is given by the following expression:
\begin{eqnarray}
[{\cal E}_{c}^{(S_{1})}, {\cal E}_{d}^{(S_{2})}] = \nonumber \\
g C_{cdf} 
\left[ \int_{s_{1} \in S_{1}} \vec{E}_{f}(s_{1}) \vec{n}_{S_{1}}(s_{1})
N(c(\vec{s}_{1}, \vec{a}); S_{2}) d \sigma \right. + \nonumber \\
+ \left. \int_{s_{2} \in S_{2}} \vec{E}_{f}(s_{2}) \vec{n}_{S_{2}}(s_{2})
N(c(\vec{s}_{2}, \vec{a}); S_{1}) d \sigma \right]
\label{t23}
\end{eqnarray}
where $N$'s are the net numbers of piercings:
\begin{eqnarray}
N(c(\vec{s}_{1}, \vec{a}); S_{2}) 
= \sum_{t_{i}(s_{1})} sgn \left(
\frac{ \partial \vec{c}(\vec{s}_{1}, \vec{a}, t_{3})}{\partial t_{3}} \cdot
\vec{n}_{S_{2}}(t_{1}, t_{2}) \right)
\label{t24}
\end{eqnarray}
with $t_{i}(s_{1})$ being solutions of the following equation:
\begin{equation}
\vec{c}(\vec{s}_{1}, \vec{a}, t_{3}) = \vec{s}_{2}(t_{1}, t_{2})
\label{t25}
\end{equation}
Eqn (\ref{t25}) is fulfilled whenever, for a given $s_{1} \in S_{1}$ the 
homotopy curve $c(\vec{s}_{1}, \vec{a})$ crosses the surface $S_{2}$.
Changing $s_{1} \rightarrow s_{2}$, $S_{1} \rightarrow S_{2}$ in eqns 
(\ref{t24}), (\ref{t25}) one gets $N$ from the second integral on the r.h.s.
of eqn(\ref{t23}). Making in (\ref{t23}) transition $S_{2} \rightarrow S_{1}$
we get for $S_{1} = S_{2} = S$:
\begin{eqnarray}
[{\cal  E}_{c}^{(S)}, {\cal E}_{d}^{(S)}]_{D} 
= 2 g c_{cdf} \int_{s \in S}
\vec{E}_{f}(s) \cdot \vec{n}_{S}(s) N(c(s, a); S) d \sigma
\label{t26}
\end{eqnarray}
In this case there is always at least one common point of $c(\vec{s}, \vec{a})$
and $S$, as $c(\vec{s}, \vec{a})$ ends on $s \in S$. The weight of this end-
point contribution to $N$ is $\frac{1}{2}$ as can be seen from the limiting 
transition $S_{1} \rightarrow S_{2}$ in eqn(\ref{t23}). Therefore, for any fixed 
$s \in S$:
\begin{eqnarray}
2 N(c(s,a); S) = sgn \frac{\partial \vec{c}(\vec{s}, \vec{a}, t_{3})}{\partial
t_{3}} |_{t_{3}=1} \cdot \vec{n}_{S}(\vec{s}) + \nonumber \\
+ 2 \sum_{t; t_{3} \ne 1, s \ne s(t_{1}, t_{2})} sgn \left( \frac{\partial
c(\vec{s}, \vec{a}, t_{3})}{\partial t_{3}} \cdot \vec{n}_{S}(t_{1}, t_{2})
\right)
\label{t27}
\end{eqnarray}
with
\begin{equation}
\vec{c}(\vec{s}, \vec{a}, t_{3}) =  \vec{s}(t_{1}, t_{2})
\label{t28}
\end{equation}
Eqns (\ref{t21}), (\ref{t26}) together with the trivial bracket
\begin{equation}
[{\cal B}_{c}, {\cal B}_{d}]_{D} = 0
\label{t29}
\end{equation}
do not form closed algebra for any chosen $H(L, c)$ and $S$. They can be
however replaced by a set of closed algebras on the properly chosen parts
of $H(L, c)$ and $S$. This will be discussed elsewhere \cite{l5}. Let us 
conclude with a choice of such $H(L_{0}, c)$ and $S_{0}$ that
\begin{equation}
L_{0} \bigcup S_{0} = \emptyset
\label{t30}
\end{equation}
i.e. abelian part does not contribute to (\ref{t21}). Moreover, put $2 N$
in eqn(\ref{t26}) and $N$ in eqn(\ref{t21}) equal to $1$. (Example: in the
Fock-Schwinger gauge take $H(L_{0}, C)$ to be a cone and $S_{0}$ to be any 
planar surface containing elliptic section of $H_{0}$). In such a case we have:
\begin{equation}
[{\cal E}_{a}^{S_{0}}, {\cal B}_{b}^{H_{0}}]_{D} = g c_{abc} {\cal B}_{c}^{H_{0}}
\label{t31}
\end{equation}
\begin{equation}
[{\cal E}_{a}^{S_{0}}, {\cal E}_{b}^{S_{0}}]_{D} = g c_{abc} {\cal E}_{c}^{S_{0}}
\label{t32}
\end{equation}
\begin{equation}
[{\cal B}_{a}^{H_{0}}, {\cal B}_{b}^{H_{0}}]_{D} = 0
\label{t33}
\end{equation}
If we took $S^{0}$ to be closed surface surrounding $\vec{a}$, then (\ref{t32}) 
is the algebra of charges  contained in its interior, $V_{0}$:
\begin{equation}
[Q_{a}^{E}(V_{0}), Q_{b}^{E}(V_{0})]_{D} = g c_{abc} Q_{c}^{E}(V_{0})
\label{t34}
\end{equation}

The question whether replacement of fluxes ${\cal B}^{H}$ by 
${\cal B}^{\sigma}$ - 
$\sigma$ being arbitrary surface - leaves simplicity of flux algebras intact 
will be discussed elsewhere \cite{l5}.

\section*{Acknowledgements}
I am indebted to dr.Gregory Korchemsky for his remark that the gauge
condition (\ref{j1}) has been already proposed and applied 
in QCD under the name
of contour gauges \cite{ll1}.
I would also like to thank professors Wojciech Kr\'{o}likowski, Elliot 
Leader 
and Mr.Jaros\l aw Boguszy\'{n}ski for interesting discussions.\\
This work was supported in part by Polish KBN Grant
PB2-P03B-06510.

\end{document}